# Mechanism of parametric pumping of magnetization precession in a nanomagnet. Parametric mechanism of current-induced magnetization reversal.

*Vadym Zayets*

*Platform Photonics Research Center (PPRC), National Institute of Advanced Industrial Science and Technology (AIST), Tsukuba, Japan*
*e-mail: v.zayets@gmail.com*

*Abstract:* A mechanism of current-induced magnetization reversal based on the parametric resonance is described. The source of the magnetization reversal is a current- induced magnetic field, which is applied perpendicularly to the easy axis of magnetic anisotropy of a ferromagnetic nanomagnet. The current- induced magnetic field was measured in a FeCoB nanomagnet to be 60 Gauss at a current density of 65 mA/μm$^2$. Two mechanisms of the magnetization reversal are described and calculated. The first mechanism is the reversal by a RF electrical current, which modulated at a frequency close to the precession frequency of the nanomagnet. The second mechanism is the reversal by a DC electrical current, in which the magneto-resistance and the current dependency of the induced magnetic field create a positive feedback loop, which amplifies a random tiny thermal fluctuation into a large magnetization precession leading to the magnetization reversal. The combination of the proposed mechanism with conventional magnetization- reversal mechanisms such as the Spin Torque and the Spin-Orbit Torque can improve the performance of a Magnetic Random Access Memory.

## *1. Introduction.*

Magnetization reversal by an electrical current[1,2] is used as a writing mechanism for a Magnetic Random Access Memory (MRAM). A cell of the MRAM memory consists of a "pin" and "free" ferromagnetic layers. The magnetic anisotropy of the "pin" layer is large and its magnetization direction is firmly fixed. The magnetic anisotropy of the "free" layer is smaller and its magnetization direction can be changed by spin injection from the "pin" layer. Data are stored in the MRAM cell by means of two stable magnetization directions of the "free" layer[3].

There are two types of MRAM design: 2-terminal and 3- terminal MRAM. The writing mechanism is different for each MRAM type. In the case of the 2-terminal MRAM[3], the writing electrical current flows through the interface between the "pin" and the "free" layers. The electrical current is spin-polarized and carries the spin-polarized conduction electrons from the "pin" layer into the "free" layer. As a result, the spin direction of spin-polarized electrons in the "free" layer is changed. When the spin injection is sufficient the magnetization of the "free" layer is reversed and the data is memorized[3-5]. This effect of magnetization reversal is called the Spin Torque (ST)[1,2].

In the case of the 3-terminal MRAM[6-8], the writing electrical current flows only inside the "pin" layer along its interface with the "free" layers. Due to the spin-dependent scattering at the interface, the spin-polarized electrons are accumulated at the interface and diffuse deep into the "free" layer. The diffused spin- polarized electrons create the spin torque, which may reverse the magnetization of the "free" layer. The spin accumulation is particularly effective at heave metal/ ferromagnetic metal interface. This effect is called the Spin- Orbit Torque (SOT)[6,7]. Avoidance of a large current flow through the tunnel junction at the interface between "pin" and "free" layers during a writing event is the main merit of the 3-termianal MRAM, which greatly improves its durability and reliability.

Both mechanisms the ST and SOT are based on the injection[8-13] of spin polarized conduction electrons from the "pin" to the "free" layer. However, the effectiveness of the spin injection substantially depends on the quality of the interface, the type of materials at the interface, and the roughness and sharpness of the interface[8-13]. As a result, the effectiveness of the magnetization reversal by the spin injection becomes very sensitive to the material choice and the fabrication technology. This severely limits possible applications of the MRAM.

This manuscript proposes and studies another mechanism of magnetization reversal, which does not rely on spin injection and, therefore, can be used with a wider range of materials and with lower requirements for the interface quality. The expected benefit is the widening of the application range of the MRAM. The principle of the proposed mechanism is based on the parametric resonance. Two cases of the magnetization reversal are described. The first case is the reversal by a RF electrical current, which modulated at a frequency close to the precession frequency of the nanomagnet. The second case is the reversal by a DC electrical current. Both cases are based on the same physical mechanism.

The source of the magnetization reversal for the described mechanism is a current- induced magnetic field, which is applied perpendicularly to the easy axis of magnetic anisotropy of a ferromagnetic nanomagnet. Under the current-induced magnetic field, the magnetization of the nanomagnet slightly inclines towards the field. Typically, the current- induced field is relatively small. It is about 60 Gauss, which is substantially smaller than the measured anisotropy field[14,15] of about 10 kGauss. The anisotropy field keeps the nanomagnet magnetization along its magnetic easy axis[14]. The magnetization inclination angle under the current induced-magnetic field is small (~200 mdeg)[14] and the magnetic field alone is not capable reversing the magnetization. However, when the electrical current and, therefore, the induced magnetic field are modulated at a frequency close to the frequency of the ferromagnetic resonance (FMR), there is a parametric enhancement of the precession and even the small current- induced oscillating magnetic field is sufficient to reverse the nanomagnet magnetization and, therefore, to memorize the data.

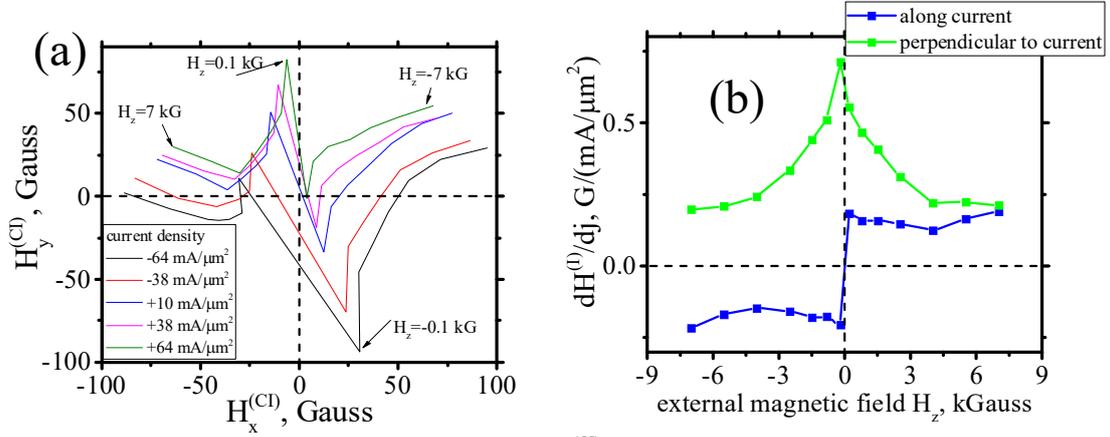

Figure 1. (a) Measured current-induced magnetic field $H^{(CI)}$ as a function of current density and external perpendicular-to-plane magnetic field $H_z$. (a) measured x- and y-components of $H^{(CI)}$. Lines of a different color correspond to a different current density. The magnetic field $H_z$ is used a parameter and scanned from -7 kGauss to +7 kGauss. (b) components of derivative $dH^{(CI)}/dj$ along and perpendicular to current.

The parametric resonance can assist the magnetization reversal not only in the case when the current is modulated at the FMR frequency, but even in the case of a DC current, when the current is not modulated at all. Such a parametric magnetization reversal by a DC current occurs in a magneto-resistant structure. In this case, a magnetization oscillation modulates the current, the current correspondingly modulates the magnetic field and the magnetic field parametrically enhances the same oscillation. There is a positive feedback loop, in which a magnetization oscillation creates a pump torque enhancing the magnitude of the same oscillation. As a result, a tiny thermally-activated magnetization oscillation amplifies itself until the magnetization reversal occurs.

The paper is organized as follows. Chapter 2 describes a measurement of current-induced magnetic field in a FeCoB nanomagnet. This field is the source of the studied mechanism of the parametric magnetic reversal. Chapter 3 describes the parametric resonance in a nanomagnet induced by a RF current and Chapter 4 describes the parametric resonance induced by a DC current. The condition of the magnetization reversal is determined by a balance between the precession pumping and the precession damping torques. In the Appendixes the pumping and damping torques are calculated for different physical mechanisms by solving the Landau–Lifshitz equation. The damping torque is calculated in Appendix 1 in the case, when a large external field is applied to a nanomagnet and when the precession damping coefficient is independent of the precession angle. The damping torque is calculated in Appendix 2 in the absence of the external magnetic field. The damping torque is calculated quantum-mechanically in Appendix 3 for the case of an interaction of magnetization with a flux of non-zero-spin particles like photons and magnons. The parametric-pumping torque is calculated in Appendix 4. All results are obtained for the case when the equilibrium magnetization is perpendicular-to-plane. The extension of the results for the in-plane equilibrium magnetization is straightforward.

## 2. *Current-induced magnetic field $H^{(CI)}$. Measurement in a FeCoB nanomagnet.*

This chapter describes the measurement of the current-induced magnetic field $H^{(CI)}$ in a FeCoB nanomagnet. The uniqueness of $H^{(CI)}$ is that an electrical current, which flows in a nanomagnet, induces a magnetic field $H^{(CI)}$, which correspondingly affects the magnetization of the same nanomagnet. The described effects of the parametric pumping of the magnetization precession and the parametric magnetization reversal exist due to this unique feature of $H^{(CI)}$.

At present, the physical origin of $H^{(CI)}$ is unknown. It was suggested[16,17] that the magnetic field $H^{(CI)}$ is originated by the spin-accumulated conduction electrons at an interface of the nanomagnet. The spin accumulation at an interface is also the source of the spin-orbit torque (SOT)[6-8,16,17]. Therefore, a measurement of $H^{(CI)}$ is used to estimate the magnitude of the spin-orbit torque[16-18]. It should be noted that dependence of the magnetic field $H^{(CI)}$ on the magnitude and the polarity of the electrical current is only a feature, which is required to excite the parametric resonance by an electrical current. The presented results are independent of the origin of the current-induced magnetic field $H^{(CI)}$.

A measurement of the current-induced magnetic field $H^{(CI)}$ in a nanomagnet is a challenging task, because of the small size of the nanomagnet. An additional measurement challenge is that the measured magnetic field $H^{(CI)}$ is very small. Typically, it is only about ~10-50 Gauss. A conventional measurement method of $H^{(CI)}$ is the method of the second harmonic[16-18], in which the current in the nanomagnet is modulated at a low frequency (~ 1 kHz) and the second harmonic of the Hall voltage is measured by the lock-in technique. The current modulates $H^{(CI)}$ and, therefore, the magnetization direction. The Hall voltage is linearly proportional to the current and to the perpendicular component of the magnetization and they both are modulated. The frequency beating between the oscillating current and magnetization direction creates the second harmonic of the Hall voltage, which is proportional to the current-induced magnetic field $H^{(CI)}$. This conventional indirect measurement method suffers from a poor measurement precision and several systematic errors[19].

Recently, a new high-precision measurement method of $H^{(CI)}$ has been proposed[19]. The merits of the proposed method are a high precision (~1 Oe), a high-reproducibility and repeatability. In the proposed method, the magnetic field $H^{(CI)}$ is measured by scanning the external magnetic field $H_{ext}$ perpendicularly to the magnetization direction and a DC measurement of the Hall voltage. $H^{(CI)}$ is evaluated from the symmetry of the scan with respect to the reversal of the $H_{ext}$ direction. The details of the measurement method are published elsewhere.

The magnetic field $H^{(CI)}$ is measured in a FeCoB (thickness is 1.1 nm) fabricated on top of a Ta (thickness is 2.5 nm) nanowire contacted by a pair of Hall probes. The equilibrium magnetization of all measured nanomagnet is perpendicular-to-plane[20]. About 100 nanomagnets, the sizes of which vary between 50 x 50 nm to 3000 x 3000 nm, were measured. The external in-plane magnetic field is scanned either along or perpendicular to the current for an individual measurement of each component of $H^{(CI)}$. Additionally, a perpendicular- to- plane magnetic field $H_z$ is applied in order to verify the dependency of the $H^{(CI)}$ on the Perpendicular Magnetic Anisotropy (PMA)[14,15].

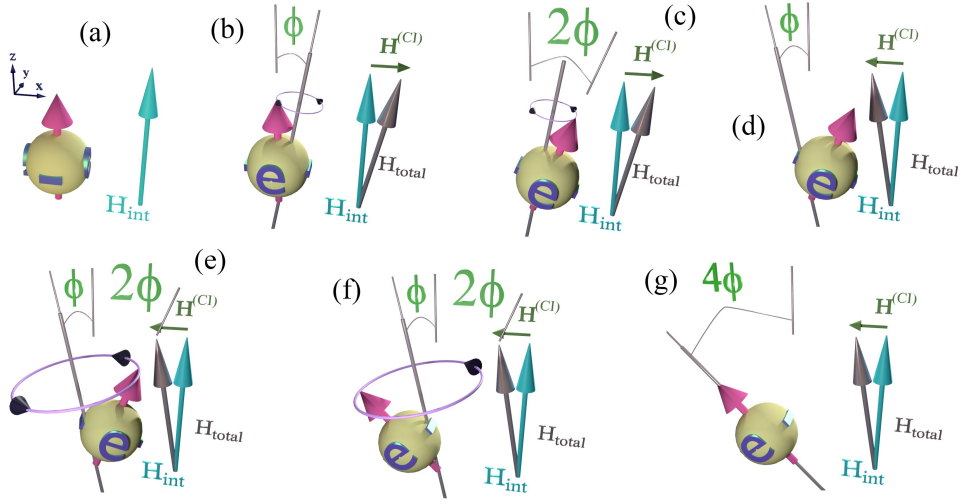

Figure 2. Parametric enhancement of magnetization oscillation under oscillating magnetic field $H^{(CI)}$ (green arrow). (a) Equilibrium, Magnetization (green ball) is directed along internal magnetic field $H_{int}$ (blue arrow). (b) When $H^{(CI)}$ is applied, the total magnetic field $H_{total}$ (grey arrow) turns away from the equilibrium direction at angle $\phi$ and magnetization precession starts; (c-f) precession angle increases due to synchronized change of $H^{(CI)}$; (g) precession angle becomes $4\phi$ after one period of precession.

Figure 1 shows the measured $H^{(CI)}$ as a function of $H_z$ and the current density for a 3000 x 3000 nm nanomagnet. The $H^{(CI)}$ depends substantially on both the $H_z$ and the current density and both dependencies are related to each other. The component of $H^{(CI)}$, which is induced by the current, is always perpendicular to the component of $H^{(CI)}$, which is induced by perpendicular magnetic field $H_z$. Since $H_z$ modulates the PMA strength[15], the observed dependence implies that the current- induced $H^{(CI)}$ is substantially influenced by the PMA effect[15].

### 3. Magnetization reversal by a RF electrical current. Enhancement of magnetization reversal by the parametric resonance.

The measured current- induced magnetic field $H^{(CI)}$ is small and an external magnetic field of the same strength is far insufficient to reverse the nanomagnet magnetization. A typical recoding current density in MRAM is about 50-100 mA/μm². At a current density of 65 mA/μm², the measured value of $H^{(CI)}$ is 60 Gauss (Fig.1). The internal magnetic field $H_{int}$, which holds the magnetization perpendicularly- to- plane, approximately equals to the anisotropy filed, which was measured to be about 10 kGauss. Applying external magnetic field of 60 Gauss perpendicularly to the 10 KGauss turns the magnetization out of the easy axis very slightly at an angle of about 200 mdeg. The turning angle is too small and does not lead to the magnetization reversal[21].

The reason why the magnetization reversal occurs under such a small magnetic field is the parametric resonance. When the electrical current is modulated at a RF frequency close to the magnetization- precession frequency, the current-induced magnetic field is modulated as well and is in a resonance with the magnetization precession. As a result, the precession is enhanced and the precession angle becomes larger after each oscillation period until the magnetization reversal occurs.

Figure 2 is the schematic diagram, which explains the mechanism of the enhancement of the magnetization precession by the parametric resonance. Figure 2(a) shows the initial states, at which the magnetization of the nanomagnet is perpendicular to the plane (along the z- axis) due to the PMA effect. There is an intrinsic magnetic field $H_{int}$, which is directed along the z- axis and which keeps the magnetization along that axis. When an electrical current flows through the nanomagnet, it creates the magnetic field $H^{(CI)}$, the direction of which is perpendicular to $H_{int}$ (along the x-axis in Fig.2b). The total magnetic field $H_{total}$, which is applied to the nanomagnet, is a vector sum of $H_{int}$ and $H^{(CI)}$. As soon as the $H^{(CI)}$ is applied and the direction of $H_{total}$ becomes different from the magnetization direction, the magnetization precession starts around the $H_{total}$. At this moment, the precession angle is the angle $\phi$ between directions of the magnetization and the $H_{total}$. As was mentioned above, the angle $\phi$ is very small, about a hundred millidegrees. After a half of the precession period, the magnetization rotates by an angle $2\phi$ with respect to the z-axis (Fig.2c). At this moment the direction of $H^{(CI)}$ is reversed and the inclination of $H_{total}$ with respect to the z- axis is reversed as well, but the absolute value of the inclination angle still remains equal to $\phi$ (Fig.2d). In total, the angle between the magnetization and the $H_{total}$ becomes $2\phi+\phi=3\phi$ (Fig.2e) and, therefore, the precession angle becomes $3\phi$. During the following precession, the angle between the magnetization and the z-axis increases and becomes $3\phi+\phi=4\phi$ after a half period of the precession (Fig.2h). This means that in case when the oscillation of the $H^{(CI)}$ is synchronized with the magnetization precession, at each precession period the precession angle increases by an angle of $4\phi$. Even though the initial field- inclination angle $\phi$ might be small, the precession angle becomes large within a short time due to the parametric pumping of the precession.

Additionally to the precession pumping, there is a precession damping, which acts in the opposite direction to the precession pumping and reduces the precession angle. The magnetization reversal occurs either when the precession pumping is larger than the precession damping at any precession angle or when the precession pumping and damping are comparable so that the precession angle can become sufficiently large for a thermally- activated magnetization switching[21].

The parametric resonance for the magnetization precession is described by the Landau-Lifshitz (LL) equation[22] as

$$\frac{\partial \vec{m}}{\partial t} = -\gamma \vec{m} \times \vec{H}_{total} - \lambda(\theta) \cdot \vec{m} \times (\vec{m} \times \vec{H}_{total}) - \gamma \vec{m} \times \vec{H}^{(CI)}(t) \quad (1)$$

where $\vec{m}$ is an unit vector directed along the magnetization, $\vec{H}_{total}$ is the total magnetic field, which is a vector sum of a intrinsic magnetic field $\vec{H}_{int}$ and an externally-applied magnetic field $\vec{H}_{ext}$; $\vec{H}^{(CI)}$ is a magnetic field, which is modulated at a frequency closed to the Larmor frequency $\omega_L$ and which excites a parametric resonance, $\gamma$ is the electron gyromagnetic ratio and $\lambda$ is the precession damping coefficient, which in general depends on the precession angle $\theta$ (See Appendixes 3 and 4). During the precession, the absolute value of the magnetization does not change. As a result, there are two independent variables: the precession angle $\theta$ and angle $\phi$ (see Fig.3b).

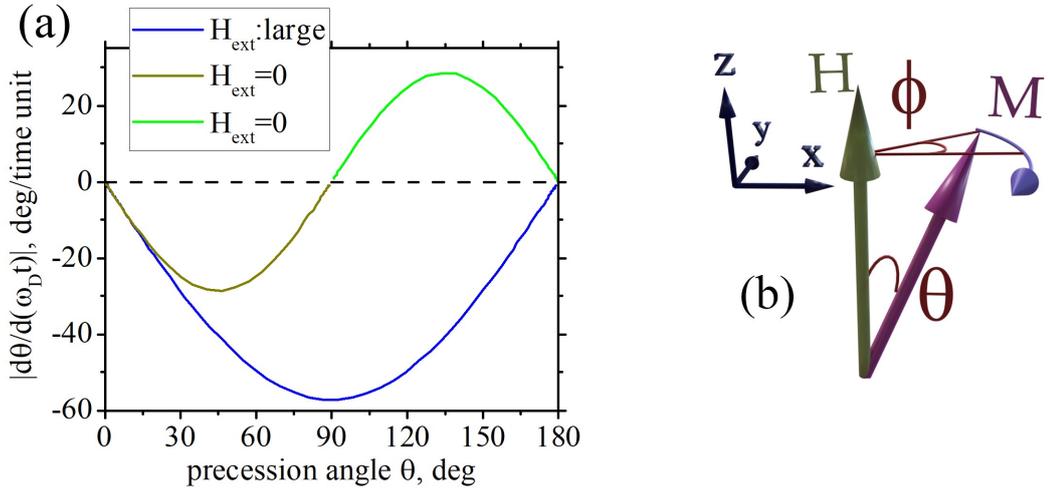

Figure 3 Precession damping torque vs. precession angle. The blue line shows case when a large external magnetic field is applied along the easy axis. The green lines show the case in the absence of an external magnetic field. The light green line shows the case when initial precession angle $\theta_0=91^0$ and the dark green line shows the case when $\theta_0=89^0$. The time is normalized to the damping rate $\omega_D$. (b) Magnetization precession around magnetic

The first left term of Eq.(1) describes the spin precession. The second and third terms describe the process of precession damping and pumping, correspondingly. The precession damping aligns the magnetization along the total magnetic field $H_{total}$ and, therefore, reduces the precession angle $\theta$. The damping torque is defined as $\left(\frac{d\theta}{dt}\right)_{damp}$ and is always negative. The damping torque can be found from LL equation (1) without the pumping term:

$$\frac{\partial \vec{m}}{\partial t} = -\gamma \vec{m} \times \vec{H}_{total} - \lambda(\theta) \cdot \vec{m} \times (\vec{m} \times \vec{H}_{total}) \quad (2)$$

The precession angle $\theta$ increases due to the precession pumping. Similarly, the damping torque is defined as $\left(\frac{d\theta}{dt}\right)_{pump}$ and is always positive. The parametric pumping torque can be found from LL equation (1) without the damping term:

$$\frac{\partial \vec{m}}{\partial t} = -\gamma \vec{m} \times \vec{H}_{total} - \gamma \vec{m} \times \vec{H}^{(CI)}(t) \quad (3)$$

A stable magnetization precession occurs when there is a balance between precession damping and pumping. A stable precession angle can be found from the condition of equal damping and pumping torques:

$$\left(\frac{d\theta}{dt}\right)_{pump} = -\left(\frac{d\theta}{dt}\right)_{damp} \quad (4)$$

The magnetization reversal occurs when the pumping torque is larger than the damping torque for any precession angle $\theta$:

$$\left(\frac{d\theta}{dt}\right)_{pump} > -\left(\frac{d\theta}{dt}\right)_{damp} \quad \theta \in \{0, \pi\} \quad (5)$$

A thermally-assisted magnetization reversal[21] may occur even at a smaller pumping torque.

In the case when the damping coefficient $\lambda$ is independent of the precession angle $\theta$, a rigorous analytical solution of Eq.(2) is expressed as (See Appendix 1):

$$m_x(t) = \cos(\omega_L t) \cdot \sin(\theta(t))$$
$$m_y(t) = \sin(\omega_L t) \cdot \sin(\theta(t)) \quad (6)$$
$$m_z(t) = \cos(\theta(t))$$

$$\theta(t) = 2 \cdot \arctan\left[e^{-\omega_D \cdot t} \tan\left(\frac{\theta_0}{2}\right)\right] \quad (7)$$

where $\omega_L = \gamma H_z$ is the Larmor frequency, $\omega_D = \lambda H_z$ is the damping rate.

The precession damping torque is calculated from Eq.(7) (See Eq.A1.10 of Appendix 1) as

$$\left(\frac{\partial \theta}{\partial t}\right)_{damp} = -\omega_D \sin(\theta) \qquad (8)$$

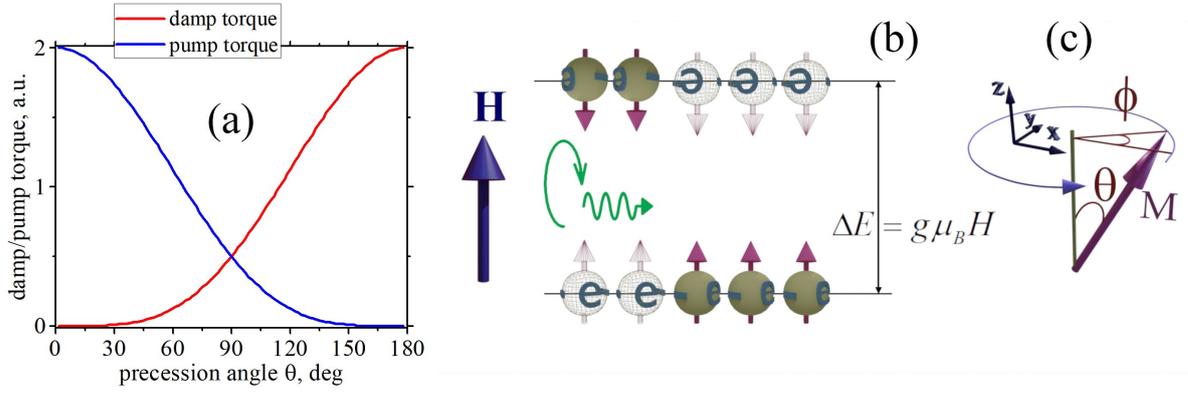

Figure 4 Precession damp/pump torque due to emission/ absorption of circularly- polarized photons. (b),(c) Two representations of the identical spin precession (b) as partially filled spin-up and spin-down energy levels; (c) as magnetization precession around magnetic field H:

The blue line of Fig.3 shows the precession damping torque, which is calculated from Eq.(8). The damping torque is small near the equilibrium at $\theta \sim 0^0$, but its magnitude monotonically increases for a larger precession angle $\theta$ until $\theta=90^0$. The damping torque is always negative. As a result of the damping, the magnetization is aligned along the external magnetic field independently of the initial precession angle $\theta_0$.

The LL Eq.(2) describes the magnetization precession around the total magnetic field $H_{total}$, which is the sum of the intrinsic magnetic field $H_{int}$ and the external applied magnetic field $H_{ext}$. Additionally, the intrinsic magnetic field is a sum of the demagnetization field $H_{demag}$ and the magnetic field[15] of the spin-orbit interaction $H_{SO}$. Both $H_{demag}$ and $H_{SO}$ are linearly proportional to the component of the magnetization $M_z$, which is perpendicular to the nanomagnet surface. When the magnetization precession angle $\theta$ increases, M inclines from the precession axis and therefore the component $M_z$ decreases. As a result, $H_{total}$ decreases. As can be seen from the LL Eq. (2), this leads to a decrease of the Larmor frequency $\omega_L$ and a decrease of the precession damping. In Appendix 2, the LL equations is solved for this case and the precession damping torque is calculated as

$$\left(\frac{\partial \theta}{\partial t}\right)_{damp} = -(\omega_{De} + \omega_{Di} \cdot \cos(\theta))\sin(\theta) \qquad (9)$$

where $\omega_{De} = \lambda H_{ext}$ and $\omega_{Di} = \lambda H_{int}$ are the damping frequencies due to the internal and external magnetic fields.

The green lines of Fig.3 show the precession damping torque, which is calculated from Eq.(9) in the absence of an external magnetic field $\omega_{De}=0$. The final alignment of the magnetization depends on the initial precession angle $\theta_0$. When $0^0<\theta<89^0$, the precession damping torque is negative and the magnetization is aligned along the z-axis $\theta=0^0$. When $90^0<\theta<180^0$, the precession damping torque is positive and the magnetization is aligned opposite to the z-axis $\theta=180^0$.

The dependence of the damping coefficient $\lambda$ in LL Eq.(2) on the precession angle may be more complex than it was described in Appendixes 1,2 and it is specific to the physical mechanism responsible for the precession damping. For example, the precession damping torque due to an emission of a circularly- polarized photon and an interaction with a magnon are different from the torques described by Eqs.(8),(9) and is calculated in Appendix 3.

The precession damping is not a spin-conserving process and requires a spin transfer from an external particle with a non-zero spin. The process of the transfer from a particle with a non-zero spin (a photon, a magnon, etc.) must be described quantum – mechanically. The magnetization precession has two absolutely identical representations. The first representation (Fig. 4c) is the rotation of the magnetization around the precession axis (the z- axis in Fig.4c), which is along the applied magnetic field H. The second representation is a sum of the spin-up and spin-down electrons (Fig. 4b), the spin direction of which is along and opposite to H. There is an energy difference between the spin-up and spin-down electrons, which is called Zeeman splitting (See Eq. A3.4). The two representations are absolutely equal and can be recalculated from each other. For example, the number of spin up $N_\uparrow$ and spin-down $N_\downarrow$ are calculated from the precession angle $\theta$ as (See Eq. (A3.12)):

$$n_\uparrow = \frac{1+\cos(\theta)}{2}$$
$$n_\downarrow = \frac{1-\cos(\theta)}{2} \qquad (10)$$

where $n_\uparrow = \frac{N_\uparrow}{N_\uparrow + N_\downarrow}$ $n_\downarrow = \frac{N_\downarrow}{N_\uparrow + N_\downarrow}$ are relative numbers of the spin up and spin-down electrons.

The spin can be transferred from the magnetization to an external particle. During this process the electron energy is reduced, the number of spin-up electrons increases and the number of the spin-down electrons decreases. This results in a decrease of the precession angle. This process is called the precession damping and the damping torque is calculated as (See Eq. (A3.17))

$$\left(\frac{\partial \theta}{\partial t}\right)_{damp} = -R_d \frac{n_\downarrow^2}{\sqrt{n_\downarrow(1-n_\downarrow)}} \qquad (11)$$

where $R_d$ is the rate of the spin transfer to the external particles.

Similarly, the spin can be transferred from a spin particle to the magnetization. During this process the electron energy increases, the number of spin-up electrons decreases and the number of the spin-down electrons increases. This results in an increase of the precession angle. This process is called the precession pumping and the pumping torque is calculated as (See Eq. (A3.18))

$$\left(\frac{\partial \theta}{\partial t}\right)_{pump} = R_p \cdot \frac{(1-n_\downarrow)^2}{\sqrt{n_\downarrow(1-n_\downarrow)}}$$

where $R_p$ is the rate of the spin transfer from the external particles.

Additionally, the rates $R_d$ and $R_p$ may also depend on the precession angle. An example is the spin transfer from/to a photon. During a magnetization precession, the magnetic moment is alternating. This means that a nanomagnet can be considered as an electro-magnetic antenna, which emits or absorbs photons. Since the spin of a circularly-polarized photon is $\pm 1$, there is a spin transfer during this process. The effectiveness of the interaction between the magnetization and a photon depends only on the alternating component of the magnetization, which is the component perpendicular to the precession axis. As a result, the rate of spin transfer can be calculated as:

$$R_d = R_{damp} \sin(\theta)$$
$$R_p = R_{pump} \sin(\theta)$$
(13)

where $R_{pump}$, $R_{pump}$ are the rates of the photon emission / absorption per unit of the alternating magnetic moment.

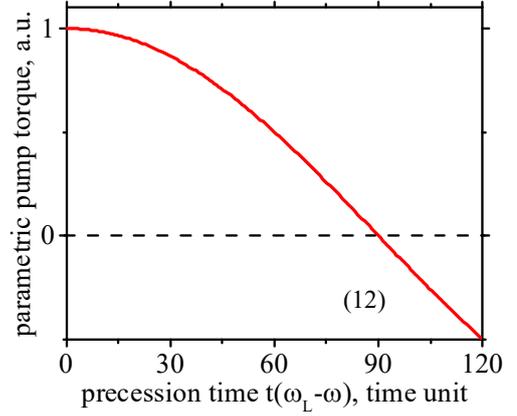

Figure 5 Temporal evolution of the precession pumping torque due to the parametric pumping by an oscillating magnetic field. The time scale is normalized to the dephasing frequency $\omega_L-\omega$. The precession oscillation and the oscillation of the magnetic field is in phase at t=0.

Substitution of Eq.(13) into Eqs.(11),(12) gives the dumping torque due a photon emission as:

$$\left(\frac{\partial \theta}{\partial t}\right)_{damp} = R_{damp} \frac{(1-\cos(\theta))^2 \sin(\theta)}{2\sqrt{1-\cos(\theta)^2}}$$
(14)

and the pumping torque due to photon absorption as:

$$\left(\frac{\partial \theta}{\partial t}\right)_{pump} = R_{pump} \cdot \frac{(1+\cos(\theta))^2 \sin(\theta)}{2\sqrt{1-\cos(\theta)^2}}$$
(15)

Figure 4 shows the spin pumping/ spin damping torque calculated from Eqs.(14),(15). In the absence of precession ($\theta=0^0$), there is no damping, but the pumping is substantial. With an increase of the precession angle, the damping monotonically increases and the pumping decreases.

Eq.(15) describes the precession pumping due to the spin transfer from a flux of external non-zero-spin particles (e.g. from photons, magnons etc.). However, it is only one of several possible mechanisms of the precession pumping. In the following, the parametric precession pumping is calculated, in which the external magnetic field is modulated at a frequency close to the Larmor frequency $\omega_L$ and the phase of the oscillating magnetic field is matched to the phase of the magnetization precession (See Fig.2).

The pumping torque for the parametric pumping can be calculated from LL Eq.(1) without the damping term:

$$\frac{\partial \vec{m}}{\partial t} = -\gamma \vec{m} \times \left(\vec{H} + \vec{H}^{(CI)}\right)$$
(16)

where H is a constant magnetic field applied along the z- axis, $H^{(CI)}$ is the oscillating magnetic field applied along the x- axis:

$$H^{(CI)} = H_x^{(CI)} \sin(\omega t)$$
(17)

The solution of Eq.(16) (See Appendix 4) gives the pumping torque for the parametric pumping as (Eq. A4.17)

$$\left(\frac{\partial \theta}{\partial t}\right)_{param} = -\gamma H_x^{(CI)} \cdot \cos\left([\omega_L - \omega]t + \varphi\right)$$
(18)

where $\phi$ is the phase between the initial Larmor oscillations and the oscillating field $H^{(CI)}$.

Figure 5 shows the temporal evolution of the parametric pump torque as calculated from Eq.(18). Even at the initial moment t=0 the parametric pumping is in phase with the phase of precession and the efficiency of the parametric pumping is largest, the dephasing between the oscillations occurs over time and the efficiency of the parametric pumping quickly decreases. Further in time the parametric precession pumping becomes zero and changes to the precession damping. From Eq.(18), the dephasing time is linearly proportional to difference between the Larmor frequency $\omega_L$ and field- oscillation frequency $\omega$. Ideally at the resonance, when $\omega = \omega_L$, there should be no dephasing and the efficiency of the parametric pumping should not decline over time. However, Eq.(18) is obtained in the absence of any precession damping. The precession damping leads to a resonance broadening, which literally means that the $\omega_L$ becomes a complex number and $\omega \neq \omega_L$ at any $\omega$. As a result, there is always a dephasing and there is always a reduction of the parametric precession pumping torque over time.

## 4. Magnetization reversal by a DC electrical current. Positive feedback loop for enlargement of precession angle.

In the previous chapter, the parametric resonance has been described for the case, when the oscillation frequency of the external magnetic field is very close to the frequency $\omega_L$ of the magnetization precession. The closeness of both frequencies is the key for the resonance enhancement of the magnetization precession. When the frequency of the external field is even- slightly detuned from the resonance condition, the dephasing occurs and the effect of the parametric precession pumping vanishes (See Fig. 5).

In this chapter another mechanism of the precession enhancement by the parametric resonance is described. In this mechanism there is no external parameter, which is modulated at the resonance frequency (the Larmor frequency). The key mechanism for this type of parametric enhancement is the positive feedback loop. The magnetization precession itself creates an oscillating magnetic field, which parametrically enhances the same magnetization precession. As a result, a small oscillation enhances itself and grows over time until the magnetization reversal occurs. Since the magnetic field of the parametric enhancement is originated by the magnetization precession itself, it is always in a perfect phase and frequency match with magnetization precession. A small thermal fluctuation seeds the process and initializes the feedback loop for the growing magnetization precession.

This effect is only possible in a magneto-resistive structure, which resistance is modulated by the magnetization precession. An example of such structure, which is shown in Fig.6, is a magnetic tunnel junction (MTJ). The magnetization is in-plane in the "pin" layer and is perpendicular to plane in the "free" layer. The magnetization of the "pin" is firmly fixed by a strong magnetic anisotropy. The magnetization precession of the "free" layer (Fig.6a) modulates the resistance of the MTJ (Fig.6b). Under a DC voltage applied to the MTJ, the electrical current is modulated at the precession frequency $\omega_L$ (Fig. 6c). The electrical current induces the magnetic field $H^{(CI)}$, which is also modulated at $\omega_L$ and, therefore, parametrically enhances the initial magnetization precession (Fig.6a).

An oscillation of the x-component of the magnetization of the "free" layer (fig.6a)

$$m_{\sim,x} = m_{\sim 0} \cdot \sin(\omega t) \qquad (19)$$

modulates the resistance R of the MTJ (Fig.6b) as:

$$R = R_0 + (R_{\uparrow\downarrow} - R_{\uparrow\uparrow})\sin(\theta)\cdot\sin(\omega t) = R_0\left[1 - k_{MR}\sin(\theta)\cdot\sin(\omega t)\right] \qquad (20)$$

where θ is the precession angle, $R_0, R_{\uparrow\downarrow}, R_{\uparrow\uparrow}$, are the MTJ when the "free" layer is perpendicular, antiparallel and parallel to the "pin" layer; $k_{MR} = (R_{\uparrow\downarrow} - R_{\uparrow\uparrow})/R_0$ is the magneto resistance.

When a DC voltage $U_{DC}$ is applied to the MTJ, the current is modulated by the resistance oscillation (Fig.6c) as

$$I = \frac{U_{DC}}{R} = \frac{I_{DC}}{1 - k_{MR}\sin(\theta)\cdot\sin(\omega t)} \qquad (21)$$

The oscillating current produces the current- induced magnetic field $H^{(CI)}$ (Fig.6d) as

$$H^{(CI)} = \varsigma I = \frac{\varsigma \cdot I_{DC}}{1 - k_{MR}\sin(\theta)\cdot\sin(\omega t)} \qquad (22)$$

If the magneto resistance is small, $k_{MR} \ll 1$, the amplitude of the oscillating current-induced field $H^{(CI)}$ can be expressed as (See Eq. (17)):

$$H_x^{(CI)} = \varsigma \cdot I_{DC} k_{MR} \sin(\theta) \qquad (23)$$

where ζ equals about 0.2-0.6 Gauss/(mA/μm²) (See Fig.1b).

The oscillating field $H^{(CI)}$ creates the parametric torque (See Eq. (18)). Since the oscillation of the $H^{(CI)}$ is exactly at the same frequency as the magnetization oscillation $\omega_L=\omega$, there is no dephasing over time $[\cos([\omega_L - \omega]t + \varphi) = 1]$ in Eq (18) and the precession pumping torque can be calculated from Eqs.(18),(23) as:

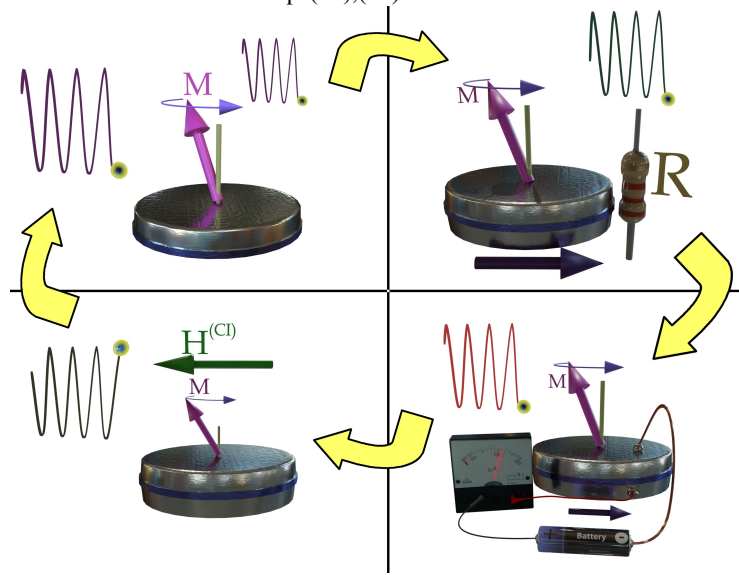

Figure 6. Positive feedback loop for an enhancement of magnetization oscillation in a nanomagnet under a DC electrical current.

$$\left(\frac{\partial \theta}{\partial t}\right)_{pump} = \gamma \cdot \varsigma \cdot I_{DC} k_{MR} \sin(\theta) \qquad (24)$$

The precession pumping torque (Eq.(24)) is linearly proportional to the magneto-resistance $k_{MR}$, the DC current $I_{DC}$ and the efficiency.

## 5. Discussion and Conclusion.

When an electrical current flows through a nanomagnet, it induces a magnetic field $H^{(CI)}$, which affects magnetization of the nanomagnet. The field is small, but measurable. At a current density of 65 mA/μm² the measured magnetic field $H^{(CI)}$ is about 60 Gauss. $H^{(CI)}$ is relatively small. For example, a static external magnetic field of the same magnitude affects the nanomagnet magnetization very weakly and its effect is barely noticeable, because the magnetization direction is kept along its easy magnetic axis by a substantially stronger internal magnetic field, which is about 10 kGauss. However, when the electrical current, which flows through the nanomagnet, and, therefore, $H^{(CI)}$ are modulated at a frequency close the resonance frequency $\omega_L$ of the magnetization precession, even the small magnetic field $H^{(CI)}$ is able to enhance the magnetization precession and even to reverse the magnetization direction. The resonance enhancement of the magnetization precession by $H^{(CI)}$ is defined as the parametric pumping. Additionally to the precession pumping, there is a precession damping, which acts in the opposite direction to precession pumping and reduces the precession angle. The effect of $H^{(CI)}$ on the nanomagnet magnetization depends on the relation between the damping and pumping torques. When at any precession angle the pumping torque is larger than the damping torque, the magnetization direction of the nanomagnet is reversed. Otherwise, there is a magnetization precession at the precession angle, at which the pumping torque is equal to the damping torque. Summing up, the dependence of both the damping torque and parametric pumping torque on the precision angle determines the features of the parametrically-induced precession and the conditions for the magnetization reversal.

Three mechanisms of the precession damping were studied. The first studied mechanism is the mechanism, for which the damping coefficient λ in LL equation is independent of the precession angle. It corresponds to the case when an external magnetic field is substantially larger than the intrinsic magnetic field in the nanomagnet. It is the simplest case for a calculation of the magnetization precession, which has a rigorous analytical solution (See Appendix 1). Even when the damping coefficient λ is independent of the precession angle θ, the calculated damping torque depends on θ (the blue line of Fig.3). The damping torque increases at a small θ and has a maximum at θ =90⁰. When the parametric pumping torque is larger than the damping torque at θ =90⁰, the magnetization is reversed.

The second studied case is the case when there is no external magnetic field. In this case the magnetization precession is around the intrinsic magnetic field of the nanomagnet. When the precession angle increases, the component of the intrinsic field along the precession axis decreases. This leads to a decrease of the precession frequency and the damping torque. The green lines of Fig.3 show the calculated damping torque in this case. The damping torque has a maximum at θ =45⁰. When the parametric pumping torque is larger than the damping torque at θ =45⁰, the magnetization is reversed.

The third studied mechanism of the precession damping is the damping due to the spin exchange between the spin of the electrons of the nanomagnet and the external flux of non-zero- spin particles like photons or magnons. The quantum- mechanical calculations indicate that this type of the precession damping is rather small when θ <30⁰ and then monolithically increases till θ =180⁰.

The parametric precession damping by the current- induced magnetic field $H^{(CI)}$ was calculated from the LL equation. The pumping torque is linearly proportional to $H^{(CI)}$. The main drawback feature of the parametric pumping is a decline of its strength over time due to the phase mismatch between the oscillating pumping field $H^{(CI)}$ and the magnetization precession. Even when at an initial moment of time the oscillations are matched, the phase mismatch is accumulated over time. The phase mismatch results in a substantial decline of the efficiency of the parametric pumping after a relatively short time (See Fig. 5).

The most unique feature of the parametric pumping is the ability to induce the magnetization precession and the magnetization reversal by a DC electrical current when there is no external parameter modulated at the precession frequency. This unique effect occurs only in a magneto-resistant structure when the electrical current is modulated by the magnetization precession and the modulated current creates the magnetic field $H^{(CI)}$, which parametrically enhances the same precession. There is a positive feedback loop, which is able to enhance the precession of a tiny thermal oscillation until a magnetization reversal occurs.

The magnetization- reversal mechanisms are very different between the studied parametric torque and the conventional torques: the Spin-Torque (ST) and the Spin-Orbit Torque (SOT). This opens an opportunity for a further optimization of the existent MRAM and an opportunity for an extension of the range of the MRAM applications. The effectiveness of the ST and SOT can be improved only by an increase of the injected amount of spin-polarized conduction electrons from the "pin" to the "free" layer. The number of possibilities for such improvement is very limited. In contrast, the parametric torque can be improved by an optimization of the positive feedback loop and by a reduction of the phase- match problem. There are much more possibilities for such an improvement.

# Appendix 1

In the following, the Landau-Lifshitz (LL) equation is solved analytically in the case when the damping term λ is independent of the precession angle. This corresponds to the case when an external magnetic field is applied along the magnetic easy axis and the external magnetic field is substantially larger than the intrinsic magnetic field (See Appendix 2 for more details).

The Landau-Lifshitz (LL) equations (1) without the pumping term can be written as

$$\frac{\partial \vec{m}}{\partial t} = -\gamma \vec{m} \times \vec{H} - \lambda \cdot \vec{m} \times \left(\vec{m} \times \vec{H}\right) \tag{A1.1}$$

where λ is independent of the magnetic field H.

Introduction of new unknowns

$$m_+ = m_x + i \cdot m_y \quad m_- = m_x - i \cdot m_y \tag{A1.2}$$

and adding/subtracting the 1st and 2nd equations of (A1.1) gives

$$\frac{\partial m_+}{\partial t} = \frac{\partial (m_x + i \cdot m_y)}{\partial t} = -\gamma H_z (m_y - i \cdot m_x) - \lambda H_z m_z (m_x + i \cdot m_y)$$

$$\frac{\partial m_-}{\partial t} = \frac{\partial (m_x - i \cdot m_y)}{\partial t} = -\gamma H_z (m_y + i \cdot m_x) - \lambda H_z m_z (m_x - i \cdot m_y) \tag{A1.3}$$

$$\frac{\partial m_z}{\partial t} = \lambda H_z \left(m_x^2 + m_y^2\right)$$

Substitution of (A1.2) into (A1.3) gives

$$\frac{\partial m_+}{\partial t} = i\omega_L m_+ - \omega_D m_z m_+$$

$$\frac{\partial m_-}{\partial t} = -i\omega_L m_- - \omega_D m_z m_- \tag{A1.4}$$

$$\frac{\partial m_z}{\partial t} = \omega_D m_+ m_-$$

where $\omega_L = \gamma H_z$ is the Larmor frequency and $\omega_D = \lambda H_z$ is the damping rate.

The solution of Eq.(A1.4) can be found as

$$\begin{pmatrix} m_+ \\ m_- \end{pmatrix} = m_{xy} \begin{pmatrix} e^{i\omega_L t} \\ e^{-i\omega_L t} \end{pmatrix} \tag{A1.5}$$

After the substitution of Eq.(A1.5) into (A1.4), the 1st and 2nd equations of (A1.4) become identical as

$$\frac{\partial m_{xy}}{\partial t} = -\omega_D m_z m_{xy} \tag{A1.6}$$

Combination of Eq.(A1.6) with the 3d equation of (A1.4) gives the system of two differential equations:

$$\frac{\partial m_{xy}}{\partial t} = -\omega_D m_z m_{xy} \tag{A1.7}$$

$$\frac{\partial m_z}{\partial t} = \omega_D m_{xy}^2$$

The solution of Eqs. (A1.7) can be found as

$$m_z(t) = \cos(\theta(t))$$
$$m_{xy}(t) = \sin(\theta(t))$$
(A1.8)

Substitution of Eq. (A1.8) into (A1.7) gives

$$\cos(\theta)\frac{\partial \theta}{\partial t} = -\omega_D \cos(\theta)\sin(\theta)$$
$$-\sin(\theta)\frac{\partial \theta}{\partial t} = \omega_D \sin(\theta)^2$$
(A1.9)

Two equations of (A1.9) are identical and can be expressed as:

$$\frac{\partial \theta}{\partial t} = -\omega_D \sin(\theta)$$
(A1.10)

Integration of Eq.(A1.10) gives

$$\int \frac{d\theta}{\sin(\theta)} = -\omega_D \cdot t + const$$
(A1.11)

The integration of Eq.(A1.11) gives the temporal evolution of the precession angle as

$$\tan\left(\frac{\theta(t)}{2}\right) = e^{-\omega_D \cdot t} \tan\left(\frac{\theta_0}{2}\right)$$
(A1.12)

where $\theta_0$ is the initial precession angle at time t=0.
From Eqs. (A1.2), (A1.5), (A1.8), the magnetization precession is described as

$$m_x(t) = \cos(\omega_L t)\cdot\sin(\theta(t))$$
$$m_y(t) = \sin(\omega_L t)\cdot\sin(\theta(t))$$
$$m_z(t) = \cos(\theta(t))$$
(A1.13)

Eq.(1.10) describes the damping torque as a function of the precession angle.

The blue line in Fig.7 shows the temporal evolution of the precession damping calculated from Eq. (A1.12). At the moment t=0, the magnetization direction is almost opposite to the direction of the magnetic field $\theta_0=179^0$. Independently of the initial precession angle, eventually the magnetization becomes parallel to the magnetic field $\theta_0=0^0$ over time.

# Appendix 2

In Appendix 1, the LL equations (Eq.2) were solved assuming that the magnetic field, which defines the precession frequency $\omega_L$ and the damping rate $\omega_D$ are independent of the precession angle θ. However, this is not a general case and $\omega_L$ and $\omega_D$ may depend on θ. The z-component of the total magnetic field is the component of the magnetic field, which defines the magnetization precession (See Fig. 3b) and, therefore, $\omega_L$ and $\omega_D$. The total magnetic field is the sum of the external and the intrinsic magnetic fields. The magnitude and direction of the intrinsic magnetic field depends on the magnetization direction and, therefore, on the precession angle. When the precession angle increases, the magnetization turns away from the precession axis and the z-component of the intrinsic field decreases. This leads to a decrease of the z- component of the total magnetic field and correspondingly to a decrease of precession frequency $\omega_L$ and the damping rate $\omega_D$.

In the following, the precession damping rate is calculated for a ferromagnetic material, in which the equilibrium magnetization direction is perpendicular-to-plane and there is no in-plane magnetic anisotropy. The origin of the Perpendicular Magnetic Anisotropy (PMA) is the spin-orbit (SO) interaction[14,15]. The contributions to the PMA from bulk atoms and the same atoms at an interface are very different[14,15], which is a feature of the SO interaction[15]. In order to simplify the calculations, it was assumed that the contributions are the same. The extension of obtained result to a realistic PMA distribution in a nanomagnet is straightforward.

In a ferromagnetic film, an electron experience 4 types of magnetic field[15]: the external magnetic field $H_{ext}$, the magnetic field induced by the magnetization $H_M$, the demagnetization field $H_{demag}$ and the magnetic field of the spin- orbit (SO) interaction[15] $H_{SO}$. The sum of the last three fields is called the intrinsic magnetic field. The demagnetization field $H_{demag}$ and the SO magnetic field $H_{SO}$ are directed perpendicularly to the film surface. $H_M$ is directed along the magnetization. The magnetization precession occurs around the z- component of the total magnetic field $H_{total}$ (See Fig3b), which can be calculated as

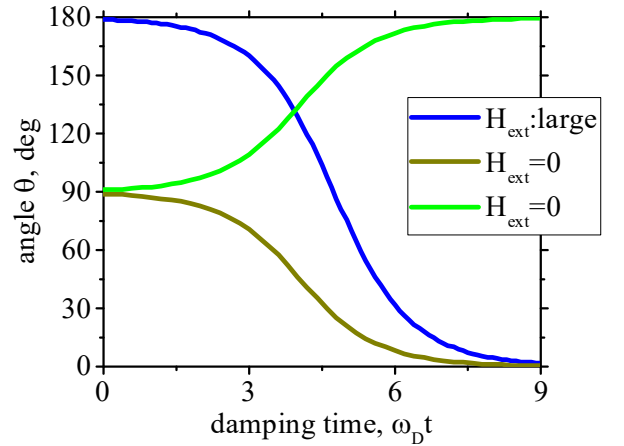

Figure 7. Temporal evolution of precession angle due to the precession damping. The blue line shows case when a large external magnetic field is applied along the easy axis. The green lines show the case in the absence of an external magnetic field. The light green line shows the case when initial precession angle $\theta_0=91^0$ and the dark green line shows the case when $\theta_0=89^0$. The time is normalized to the damping rate $\omega_D$

$$H_{total,z} = H_{ext} + M_z - k_{demag}M_z + k_{SO}M_z \quad (A2.1)$$

where $\vec{H}_M = \vec{M}$; $H_{demag} = -k_{demag}M_z$; $k_{demag}$ is the demagnetization factor[14] $k_{demag} \leq 1$; $k_{SO}$ is the SO factor[15] $H_{SO} = k_{SO}M_z$; $k_{SO} > 0$; and the external magnetic field $H_{ext}$ is applied perpendicularly to the film surface (along the z-axis).

At precession angle θ, the z-component of magnetization is calculated as

$$M_z = M \cdot \cos(\theta) \quad (A2.2)$$

where M is the total magnetization of the nanomagnet.

The intrinsic magnetic field $H_{int}$ is the sum of $H_M$, $H_{demag}$ and $H_{SO}$. The component of $H_{int}$ along the precession axis (the z-component) can be calculated from Eq. (A2.1) as

$$H_{total,z} = H_{ext} + H_{int,0} \cdot \cos(\theta) \quad (A2.3)$$

where $H_{int,0}$ is the equilibrium intrinsic magnetic field in the absence of the magnetization precession.

The solution procedure of the LL equation with $H_{total,z}$ as the magnetic field is similar to that described in Appendix 1. Substitution of the field (A2.3) into Eq.(2) and use of new unknowns $m_+$ and $m_+$ (A1.2) give

$$\frac{\partial m_+}{\partial t} = i \cdot m_+ \left(\omega_{Le} + \omega_{Li} \cos(\theta)\right) - m_z m_+ \left(\omega_{De} + \omega_{Di} \cos(\theta)\right)$$

$$\frac{\partial m_-}{\partial t} = -i \cdot m_- \left(\omega_{Le} + \omega_{Li} \cos(\theta)\right) - m_z m_- \left(\omega_{De} + \omega_{Di} \cos(\theta)\right) \quad (A2.4)$$

$$\frac{\partial m_z}{\partial t} = m_+ m_- \left(\omega_{De} + \omega_{Di} \cos(\theta)\right)$$

where the Larmor frequency and the damping rate due to the intrinsic and extrinsic magnetic fields are defined as

$$\omega_{Li} = \gamma H_{int,0} \quad \omega_{Le} = \gamma H_{ext} \quad \omega_{Di} = \lambda H_{int,0} \quad \omega_{De} = \lambda H_{ext} \quad (A2.5)$$

The solution of Eq.(A2.4) can be found as

$$\begin{pmatrix} m_+ \\ m_- \end{pmatrix} = m_{xy} \begin{pmatrix} e^{i(\omega_{Le} + \omega_{Li} \cdot \cos(\theta_M))t + i \cdot \phi(t)} \\ e^{-i(\omega_{Le} + \omega_{Li} \cdot \cos(\theta_M))t - i \cdot \phi(t)} \end{pmatrix} \quad (A2.6)$$

where $\phi(t)$ is the time-dependent phase of the precession.

Substitution of (A2.6) into (A2.4) gives

$$\frac{\partial m_{xy}}{\partial t} + i m_{xy} \left(-\omega_{Li} \sin(\theta)\frac{\partial \theta}{\partial t} \cdot t + \phi(t)\right) = -\left(\omega_{De} + \omega_{Di} \cdot \cos(\theta)\right) m_z m_{xy}$$

$$\frac{\partial m_{xy}}{\partial t} - i m_{xy} \left(-\omega_{Li} \sin(\theta)\frac{\partial \theta}{\partial t} \cdot t + \phi(t)\right) = -\left(\omega_{De} + \omega_{Di} \cdot \cos(\theta)\right) m_z m_{xy} \quad (A2.7)$$

$$\frac{\partial m_z}{\partial t} = \left(\omega_{De} + \omega_{Di} \cdot \cos(\theta)\right) m_{xy}^2$$

The complex part of Eqs. (A2.7) gives the equation for the precession phase as

$$\phi(t) = \omega_{Li} \sin(\theta)\frac{\partial \theta}{\partial t} \cdot t \quad (A2.8)$$

The real part of Eqs. (A2.7) are

$$\frac{\partial m_{xy}}{\partial t} = -\left(\omega_{De} + \omega_{Di} \cdot \cos(\theta)\right) m_{xy} m_z \quad (A2.9)$$

$$\frac{\partial m_z}{\partial t} = \left(\omega_{De} + \omega_{Di} \cdot \cos(\theta)\right) m_{xy}^2$$

The solution of Eqs. (A2.9) can be found as

$$m_z(t) = \cos(\theta(t)) \quad (A2.10)$$
$$m_{xy}(t) = \sin(\theta(t))$$

Substitution of Eq. (A2.10) into (A2.9) gives

$$\cos(\theta)\frac{\partial \theta}{\partial t} = -\left(\omega_{De} + \omega_{Di} \cdot \cos(\theta)\right)\cos(\theta)\sin(\theta) \quad (A2.11)$$

$$-\sin(\theta)\frac{\partial \theta}{\partial t} = \left(\omega_{De} + \omega_{Di} \cdot \cos(\theta)\right)\sin^2(\theta)$$

Two equations of (A2.11) are identical and can be expressed as:

$$\frac{\partial \theta}{\partial t} = -\left(\omega_{De} + \omega_{Di} \cdot \cos(\theta)\right)\sin(\theta) \quad (A2.12)$$

Integration of Eq.(2.12) gives

$$\int \frac{d\theta}{\left(\omega_{De} + \omega_{Di} \cdot \cos(\theta)\right)\sin(\theta)} = -t + const \quad (A2.13)$$

In absence of magnetic field, the integration of the left part gives

$$\tan(\theta) = e^{-\omega_{Di} t} \tan(\theta_0) \quad (A2.14)$$

where $\theta_0$ is the precession angle at t=0.

Eq.(2.12) describes the damping torque as a function of the precession angle. The green lines in Fig.7 show the temporal evolution of the precession damping calculated from Eq. (A2.14). Since at $\theta_0=90^0$ there is no damping torque, the data is calculated separately for two regions. For region $91^0<\theta<180^0$ (light-green line), the data is calculated using $\theta_0=91^0$. For region $0^0<\theta<89^0$ (dark-green line), the data is calculated using $\theta_0=89^0$. The final alignment of the magnetization depends on the initial precession angle $\theta_0$. The magnetization is aligned along the z-axis $\theta=0^0$, when the initial precession angle is smaller than $90^0$. The magnetization is aligned opposite to the z-axis $\theta=180^0$, when the initial precession angle is greater than $90^0$.

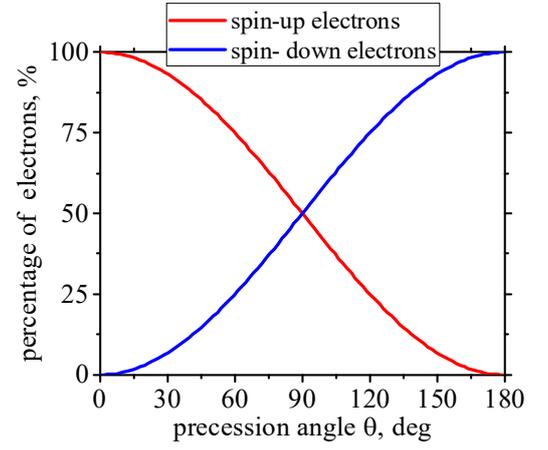

Figure 8. Percentage of spin-up and spin-down electrons as a function of the precession angle

## Appendix 3

In Appendices 1 and 2 the damping torque has been calculated based on an assumption that the damping term of the LL equations (1) is linearly proportional to the total magnetic field inside of a ferromagnetic nanomagnet. It is only an assumption, which may be not valid for each specific precession damping mechanism.

The precession damping is not a spin-conserving process and, therefore, participation of another particle with a non-zero spin (e.g. a photon, magnon etc.) is required for the precession damping to occur. In the following, the spin damping torque is calculated from a known rate of interaction of non-zero-spin particles with the magnetization.

The energy of an electron (a spin-up electron), which spin is aligned along a magnetic field H, is smaller than the energy of an electron (a spin-down electron), which spin is aligned opposite to H (See Fig.5). The energy difference is called the Zeeman energy. The magnetization precession can be represented as a quantum state, in which both the spin-down and spin-up states are partially occupied.

Absorption of a non-zero-spin particle (e.g. a photon) excites a spin-up electron to the spin-down energy level. A larger relative number of spin-down electrons corresponds to a larger precession angle $\theta$. Therefore, the absorption of a non-zero-spin particle causes an increase of the precession angle. Similarly, the emission of a non-zero-spin particle reduces the number of spin-down electrons and, therefore, reduces the precession angle $\theta$.

In the following, the dependence of precession angle on the number of spin-up and spin-down electrons is calculated using the spinor technique. A quantum state of an electron, which spin is aligned at angles $\theta$ and $\phi$ as shown in Fig.5, is described by a spinor S, which is an eigenvector of the following Pauli matrix:

$$\left[\sigma_x \cos(\phi)+\sigma_y \sin(\phi)\right]\sin(\theta)+\sigma_z \cos(\theta)=$$
$$=\left[\begin{pmatrix} 0 & 1 \\ 1 & 0 \end{pmatrix}\cos(\phi)+\begin{pmatrix} 0 & -i \\ i & 0 \end{pmatrix}\sin(\phi)\right]\sin(\theta)+\begin{pmatrix} 1 & 0 \\ 0 & -1 \end{pmatrix}\cos(\theta) \quad (A3.1)$$

The spinor S, which is calculated as an eigenvector of (A3.1), is

$$\hat{S}_{\theta,\phi}=\frac{1}{\sqrt{2(1+\cos(\theta))}}\begin{pmatrix} 1+\cos(\theta) \\ \sin(\theta)\cdot e^{-i\phi} \end{pmatrix} \quad (A3.2)$$

The wavefunction of an electron, which is described by the spinor (A3.2), can be expressed as linear combination of wavefunctions, which correspond to the spin-up ($\theta=0$) and spin-down ($\theta=\pi$) states or as a vector dot product:

$$\Psi_{\theta,\phi}=\hat{S}_{\theta,\phi}\cdot\begin{pmatrix} \Psi_\uparrow & \Psi_\downarrow \end{pmatrix} \quad (A3.3)$$

The energy difference $\Delta E_{Zeeman}$ between the electron states of the spin directions along and opposite to the magnetic field H is calculated as:

$$\Delta E_{Zeeman}=g\cdot\mu_B H \quad (A3.4)$$

where g is the g-factor and $\mu_B$ is the Bohr magneton.

The wavefunction of the spin-up and spin-down electron states can be expressed as

$$\Psi_\uparrow(\vec{r},t)=\Psi(\vec{r})\cdot e^{i\frac{E_0}{\hbar}t} e^{+i\frac{g\cdot\mu_B H}{2\hbar}t}$$
$$\Psi_\downarrow(\vec{r},t)=\Psi(\vec{r})\cdot e^{i\frac{E_0}{\hbar}t} e^{-i\frac{g\cdot\mu_B H}{2\hbar}t} \quad (A3.5)$$

Wavefunctions of (A3.5) can be expressed using the spinor representation (A3.3) as

$$\Psi_\uparrow=\begin{pmatrix} 1 \\ 0 \end{pmatrix}\cdot\begin{pmatrix} \Psi_\uparrow & \Psi_\downarrow \end{pmatrix}\cdot e^{i\left(\frac{E_0}{\hbar}+\frac{g\cdot\mu_B H}{2\hbar}\right)t}$$
$$\Psi_\downarrow=\begin{pmatrix} 0 \\ e^{-i\frac{g\cdot\mu_B H}{\hbar}t} \end{pmatrix}\cdot\begin{pmatrix} \Psi_\uparrow & \Psi_\downarrow \end{pmatrix}\cdot e^{i\left(\frac{E_0}{\hbar}+\frac{g\cdot\mu_B H}{2\hbar}\right)t} \quad (A3.6)$$

where the spin-up state corresponds to θ=0⁰ and the spin-down state corresponds to θ=180⁰. Comparison of Eqs. (A3.6) with Eq. (A3.3) gives the expression for the spinor at an arbitrary angle θ as

$$\hat{S}_{\theta,\phi} = \frac{1}{\sqrt{2(1+\cos(\theta))}} \begin{pmatrix} 1+\cos(\theta) \\ \sin(\theta) \cdot e^{-i\frac{g \cdot \mu_B H}{\hbar}t} \end{pmatrix} \quad (A3.7)$$

Eq.(A3.7) describes a spin precession at a precession angle θ and with the Larmor frequency $\omega_L$:

$$\omega_L = \frac{g \cdot \mu_B H}{\hbar} \quad (A3.8)$$

It means that the precession term of the LL equation (Eq.1) describes the Zeeman splitting and in general is a feature of the time-inverse symmetry.

The spin precession around a magnetic field can be described by a relative number of the spin-up and spin down electrons. In a simplified case when the wavefunction of the system of electrons with the spin S can be described as a sum of wavefunctions for spin-up and spin-down electrons as

$$\Psi_S = \Psi_\uparrow \cdot \sqrt{N_\uparrow} + \Psi_\downarrow \cdot \sqrt{N_\downarrow} \quad (A3.9)$$

where $N_\uparrow$ and $N_\downarrow$ are the numbers of the spin- up and spin- down electrons. The expression (A3.9) corresponds to the spinor:

$$\hat{S} = \frac{1}{\sqrt{N_\uparrow + N_\downarrow}} \begin{pmatrix} \sqrt{N_\uparrow} \\ \sqrt{N_\downarrow} \end{pmatrix} \quad (A3.10)$$

Comparison of Eqs. (A3.7) and (A3.10) gives

$$\sqrt{n_\uparrow} = \frac{1+\cos(\theta)}{\sqrt{2(1+\cos(\theta))}}$$
$$\sqrt{n_\downarrow} = \frac{\sin(\theta)}{\sqrt{2(1+\cos(\theta))}} \quad (A3.11)$$

where $n_\uparrow$ and $n_\downarrow$ are the relative numbers of the spin up and spin down electrons

$$n_\uparrow = \frac{N_\uparrow}{N_\uparrow + N_\downarrow} \quad n_\downarrow = \frac{N_\downarrow}{N_\uparrow + N_\downarrow} \quad (A3.12)$$

Simplification of Eq.(A3.11) gives the number of spin-up and spin-down electrons, when there is a magnetization precession and the precession angle is θ, as

$$n_\uparrow = \frac{1+\cos(\theta)}{2}$$
$$n_\downarrow = \frac{1-\cos(\theta)}{2} \quad (A3.13)$$

Figure 8 shows the filling percentage of the spin- up and spin-down levels as a function of the precession angle. The filling amounts are the same at a precession angle of 90⁰.

The probability $P_{pump}$ to excite one electron from the spin-up to the spin-down level and, therefore, to increase the precession angle is linearly proportional to the number of the spin-up electrons and the number of available spin-down quantum states.

$$P_{pump} = P_{pump,0} \cdot n_\uparrow \cdot (1-n_\downarrow) = P_{pump,0} \cdot (1-n_\downarrow)^2 \quad (A3.14)$$

where $P_{pump,0}$ is the probability to excite a spin-up electron into one already-empty spin-down state.

Similarly, the probability that an electron returns back to the spin-up level is calculated as

$$P_{damp} = P_{damp,0} \cdot n_\downarrow \cdot (1-n_\uparrow) = P_{damp,0} \cdot n_\downarrow^2 \quad (A3.15)$$

where $P_{damp,0}$ is the transition probability of a spin-down electron into an empty spin-up state. The $P_{pump,0}$ and $P_{damp,0}$ are defined by the interaction mechanism of the total spin of the localized electrons with external spin particles (magnons, phonons, spin-polarized conduction electrons).

The electron transition rate is defined as the transition probability per a unit of time. The rates of transitions between the spin-up and the spin- down levels or the pumping and damping rates can be calculated from probabilities (A3.14), (A3.15) as

$$\left(\frac{\partial n_\uparrow}{\partial t}\right)_{damp} = R_d \cdot n_\downarrow^2$$
$$\left(\frac{\partial n_\downarrow}{\partial t}\right)_{pump} = R_p \cdot (1-n_\downarrow)^2 \quad (A3.16)$$

where $R_d$ and $R_p$ are the rates of the spin transition from the spin-down to the spin-up level and from the spin-down to the spin-up level, correspondingly.

The spin damping torque is calculated from (A3.13) (A3.16) as

$$\left(\frac{\partial \theta}{\partial t}\right)_{damp} = \frac{\partial \theta}{\partial n_\downarrow}\left(\frac{\partial n_\downarrow}{\partial t}\right)_{damp} = -\frac{\partial \theta}{\partial n_\downarrow} R_d \cdot n_\downarrow^2 = -R_d \frac{n_\downarrow^2}{\sqrt{n_\downarrow(1-n_\downarrow)}} \quad (A3.17)$$

Similarly, the spin pumping torque is calculated as

$$\left(\frac{\partial \theta}{\partial t}\right)_{pump} = \frac{\partial \theta}{\partial n_\downarrow}\left(\frac{\partial n_\downarrow}{\partial t}\right)_{pump} = -\frac{\partial \theta}{\partial n_\downarrow} R_p \cdot (1-n_\downarrow)^2 = R_p \cdot \frac{(1-n_\downarrow)^2}{\sqrt{n_\downarrow(1-n_\downarrow)}} \qquad (A3.18)$$

The interaction of a spin particle with the magnetization of a nanomagnet also depends on the precession angle θ. For example, the emission or absorption of a circularly- polarized photon is linearly proportional to the alternating part of the magnetic moment. It is because the nanomagnet can be considered as an electro-magnetic antenna, which emits or absorbs photons, and the effectiveness of the photon absorption/ emission is proportional only to the alternating component of the magnetic moment. In the case of the precession around the z-axis, the alternating magnetic moment is promotional to the xy-component of the magnetization M (See Fig.5) and the spin pumping torque can be calculated from (A3.18) as:

$$\left(\frac{\partial \theta}{\partial t}\right)_{pump} = R_{pump} \cdot \frac{(1-n_\downarrow)^2}{\sqrt{n_\downarrow(1-n_\downarrow)}} \sin(\theta) \qquad (A3.19)$$

where $R_{pump}$ is the rate of the photon absorption per unit of the alternating magnetic moment. The $R_{pump}$ is linearly proportional to the photon flux, which irradiates the nanomagnet. The substitution of Eq.(A3. 13) into A(3.19) gives the precession pumping torque due to absorption of circular-polarized photons as

$$\left(\frac{\partial \theta}{\partial t}\right)_{pump} = R_{pump} \cdot \frac{(1+\cos(\theta))^2 \sin(\theta)}{2\sqrt{1-\cos(\theta)^2}} \qquad (A3.20)$$

Similarly, the emission of photons is also linearly proportional to the alternating magnetic moment and therefore the xy-component of the magnetization. The substitution of Eq.(A3. 13) into (A3.17) gives the precession damping torque due to emission of circular-polarized photons as

$$\left(\frac{\partial \theta}{\partial t}\right)_{damp} = -R_{damp} \frac{n_\downarrow^2}{\sqrt{n_\downarrow(1-n_\downarrow)}} \sin(\theta) = R_{damp} \frac{(1-\cos(\theta))^2 \sin(\theta)}{2\sqrt{1-\cos(\theta)^2}} \qquad (A3.21)$$

# Appendix 4

In the following, the Landau-Lifshitz (LL) equation is solved in the case of a parametric pumping by an external magnetic field, which oscillates at a frequency close to the FMR frequency. The aim of the calculation is to evaluate the pumping torque due to the parametric pumping. In order to make calculation simpler, the LL equation is solved without the damping term. The LL equation with the precession and parametric term and without the damping term is

$$\frac{\partial \vec{m}}{\partial t} = -\gamma \vec{m} \times \left(\vec{H} + \vec{H}^{(CI)}\right) \qquad (A4.1)$$

where H is applied along the z- axis, $H^{(CI)}$ is applied along the x- axis and is oscillating with the frequency ω:
$$H^{(CI)} = H_x^{(CI)} \sin(\omega t) \qquad (A4.2)$$

The scalar form of the Eq. (A4.1) is

$$\frac{\partial m_x}{\partial t} = -\omega_L m_y$$
$$\frac{\partial m_y}{\partial t} = \omega_L m_x - \omega_\sim \cdot m_z \cdot \sin(\omega t) \qquad (A4.3)$$
$$\frac{\partial m_z}{\partial t} = \omega_\sim \cdot m_y \cdot \sin(\omega t)$$

where the Larmor frequency $\omega_L = \gamma H_z$ and $\omega_\sim = \gamma H_x^{(CI)}$.

Introduction of new unknowns
$$m_+ = m_x + i \cdot m_y \quad m_- = m_x - i \cdot m_y \qquad (A4.4)$$

and addition/ subtraction of the 1$^{st}$ and 2$^{nd}$ equations of (A4.3) give

$$\frac{\partial m_+}{\partial t} = i\omega_L m_+ - i \cdot \omega_\sim \cdot m_z \cdot \sin(\omega t)$$
$$\frac{\partial m_-}{\partial t} = -i\omega_L m_- + i \cdot \omega_\sim \cdot m_z \cdot \sin(\omega t) \qquad (A4.5)$$
$$\frac{\partial m_z}{\partial t} = \omega_\sim \frac{m_+ - m_-}{2i} \cdot \sin(\omega t)$$

The solution of Eq.(A4.5) can be found as

$$m_+(t) = M \cdot \sin(\theta(t)) \cdot e^{i\phi_+(t)}$$
$$m_-(t) = M \cdot \sin(\theta(t)) \cdot e^{-i\phi_-(t)} \qquad (A4.6)$$
$$m_z(t) = M \cdot \cos(\theta(t))$$

where θ, $\phi_+$ and $\phi_-$ are new unknowns and M is the magnetization.
Substitution of (A4.6) into (A4.5) gives

$$\frac{\partial \theta}{\partial t} + i \cdot \tan(\theta) \left[ \frac{\partial \phi_+}{\partial t} - \omega_L \right] = -i \cdot \omega_\sim \cdot \sin(\omega t) \cdot e^{-i\phi_+}$$

$$\frac{\partial \theta}{\partial t} - i \cdot \tan(\theta) \left[ \frac{\partial \phi_-}{\partial t} - \omega_L \right] = +i \cdot \omega_\sim \cdot \sin(\omega t) e^{i\phi_-}$$

$$\frac{\partial \theta}{\partial t} = -\omega_\sim \frac{e^{i\phi_+} - e^{-i\phi_-}}{2i} \cdot \sin(\omega t)$$

(A4.7)

Substitution of the 3$^{rd}$ Eq. into the 1$^{st}$ and 2$^{nd}$ Eqs. of (A4.7) gives

$$\tan(\theta) \left[ \frac{\partial \phi_+}{\partial t} - \omega_L \right] = \omega_\sim \cdot \left( -e^{-i\phi_+} - \frac{e^{i\phi_+} - e^{-i\phi_-}}{2} \right) \cdot \sin(\omega t)$$

$$\tan(\theta) \left[ \frac{\partial \phi_-}{\partial t} - \omega_L \right] = \omega_\sim \cdot \left( -e^{i\phi_-} + \frac{e^{i\phi_+} - e^{-i\phi_-}}{2} \right) \cdot \sin(\omega t)$$

$$\frac{\partial \theta}{\partial t} = -\omega_\sim \frac{e^{i\phi_+} - e^{-i\phi_-}}{2i} \cdot \sin(\omega t)$$

(A4.8)

Taking into account that the oscillating field is small

$$H_{x,\omega} \ll H \quad \omega_\omega \ll \omega_L$$

(A4.9)

The solution can be expanded into a Taylor series, where $\omega_\sim$ is as a small parameter:

$$\phi_+ = \phi_{+,0} + \omega_\sim \cdot \phi_{+,1}$$

$$\phi_- = \phi_{-,0} + \omega_\sim \cdot \phi_{-,1}$$

$$\theta = \theta_0 + \omega_\sim \cdot \theta_1$$

(A4.10)

Substitution of (A4.10) into (A4.8) gives

$$\left[ \tan(\theta_0) + \omega_\sim \cdot \theta_1 + \omega_\sim \cdot \theta_1 \cdot \tan(\theta_0)^2 \right] \left[ \frac{\partial \phi_{+,0}}{\partial t} - \omega_L + \omega_\sim \frac{\partial \phi_{+,1}}{\partial t} \right] = \omega_\sim \cdot \left( -e^{-i\phi_{+,0}} - \frac{e^{i\phi_{+,0}} - e^{-i\phi_{-,0}}}{2} \right) \cdot \sin(\omega t)$$

$$\left[ \tan(\theta_0) + \omega_\sim \cdot \theta_1 + \omega_\sim \cdot \theta_1 \cdot \tan(\theta_0)^2 \right] \left[ \frac{\partial \phi_{-,0}}{\partial t} - \omega_L + \omega_\sim \frac{\partial \phi_{-,1}}{\partial t} \right] = \omega_\sim \cdot \left( -e^{i\phi_{-,0}} + \frac{e^{i\phi_{+,0}} - e^{-i\phi_{-,0}}}{2} \right) \cdot \sin(\omega t)$$

$$\frac{\partial \theta_0}{\partial t} + \omega_\sim \frac{\partial \theta_1}{\partial t} = -\omega_\sim \frac{e^{i\phi_{+,0}} - e^{-i\phi_{-,0}}}{2i} \cdot \sin(\omega t)$$

(A4.11)

Comparison of parts, which is proportional to $(\omega_\sim)^0$ gives

$$\tan(\theta_0) \left[ \frac{\partial \phi_{0,+}}{\partial t} - \omega_L \right] = 0$$

$$\tan(\theta_0) \left[ \frac{\partial \phi_{0,-}}{\partial t} - \omega_L \right] = 0$$

$$\frac{\partial \theta_0}{\partial t} = 0$$

(A4.12)

The solution of Eqs.(A4.12)) gives a static precession at the Larmor frequency $\omega_L$ and at constant precession angle $\theta_0$ as

$$\phi_{+,0} = \omega_L t + \varphi$$

$$\phi_{-,0} = \omega_L t + \varphi$$

$$\theta_0 = const$$

(A4.13)

where φ describes the phase of the oscillation with respect to the phase of the parametric pumping. This zero- approximation describes the case of the magnetization precession without the parametric pumping. Substitution of (A4.13) into (A4.4), (A4.6) gives

$$m_x(t) = M \cdot \sin(\theta_0) \cdot \cos(\omega_L t + \varphi)$$

$$m_-(t) = M \cdot \sin(\theta_0) \cdot \sin(\omega_L t + \varphi)$$

$$m_z(t) = M \cdot \cos(\theta_0)$$

(A4.14)

Comparison of parts in Eq.(A4.11), which are proportional to $(\omega_\sim)^1$, gives

$$\tan(\theta_0) \frac{\partial \phi_{+,1}}{\partial t} = \left( -e^{-i(\omega_L t + \varphi)} - \cos(\omega_L t + \varphi) \right) \cdot \sin(\omega t)$$

$$\tan(\theta_0) \frac{\partial \phi_{+,1}}{\partial t} = \left( -e^{i(\omega_L t + \varphi)} + \cos(\omega_L t + \varphi) \right) \cdot \sin(\omega t)$$

$$\frac{\partial \theta_1}{\partial t} = -\sin(\omega_L t + \varphi) \cdot \sin(\omega t)$$

(A4.15)

The solution of the third equation of Eqs.(A4.15) is

$$2\theta_1 = -\frac{\sin([\omega_L - \omega]t + \varphi)}{\omega_L - \omega} + \frac{\sin([\omega_L + \omega]t + \varphi)}{\omega_L + \omega}$$

(A4.16)

Ignoring the fast oscillating part at the frequency ω+ω_L, substitution of (A4.16) into (A4.10) gives the temporal evolution of the precession angle as

$$\theta = \theta_0 - \frac{\omega_\sim}{2} \cdot \frac{\sin\left(\left[\omega - \omega_L\right]t + \varphi\right)}{\omega - \omega_L} \tag{A4.17}$$

Substitution of the 3$^{rd}$ Eq. of (A4.15) into (A4.10) gives the pumping torque as

$$\frac{\partial \theta}{\partial t} = -\omega_\sim \cdot \cos\left(\left[\omega_L - \omega\right]t + \varphi\right) \tag{A4.18}$$